# Fast recognition of single molecules based on single event photon statistics

Shuangli Dong, Tao Huang, Yuan Liu, Jun Wang,
Guofeng Zhang, Liantuan Xiao[+], and Suotang Jia
State Key Laboratory of Quantum Optics and Quantum Optics Devices,
College of Physics and Electronics Engineering, Shanxi University, Taiyuan,
030006, China

Mandel's $Q$-parameter, which is determined from single event photon statistics, provides an alternative to differentiate single-molecule with fluorescence detection. In this work, by using the $Q$-parameter of the sample fluorescence compared to that of an ideal double-molecule system with the same average photon number，we present a novel and fast approach for identifying single molecules based on single event photon statistics analyses, compared with commonly used two-time correlation measurements. The error estimates for critical values of photon statistics are also presented for single-molecule determination.

## I. INTRODUCTION

Although people most often think about and model molecule systems in terms of individuals, experimental science has been dominated by measurements that result in ensemble averages. This has traditionally hidden much of the rich variety present at microscopic scales. Detecting single molecules optically was an old dream, already pursued by Jean Perrin at the beginning of the 20th century [1]. In the 1980s, advances in scanning tunneling microscopy (STM) and atomic force microscopy (AFM) have allowed the visualization and even manipulation of single atoms and molecules [2,3]. Optical spectroscopy offers a wealth of information on the structure, interaction, and dynamics of molecule species. The first convincing detection of a single molecule was achieved in 1989 by Moerner and Kador in an absorption measurement [4]. Soon here after, fluorescence was shown to provide a much better signal-to-noise ratio, in cryogenic condition [5] as well as at room temperature [6, 7]. The microscopy of single molecules at room temperature took off in 1993 with Betzig and Chichester's detection of immobilized molecules on a solid surface by means of excitation with a near-field optical source [8]. The scope of the method expanded suddenly when several groups [9-11] showed that single molecules could be detected at ambient conditions with a simple confocal microscope. Single-molecule microscopy by fluorescence at room temperature has now become a versatile and general technique, opening investigations of the nanoworld [12,13]. In above experiments it is critical to ascertain that the observed signal actually comes from a single molecule, but not the random mixture of emissions from nearby molecules [14]. Usually low concentration and equivalently small excitation volumes are the most common experimental strategies [15].

Once experimental conditions favoring single-molecule detection are satisfied, a number of criteria have to be met to ascertain that the observed signal actually comes from a single emitter [16, 17]. In previous experiments these criteria are all direct consequences of common photophysical properties of fluorophores. For instance, observed fluorescence intensity level is consistent with that of a single emitting molecule, fluorescence emission exhibits only *two levels* [18] and the emitted light exhibits *antibunching* [19, 20]. The most common criteria is to detect an antibunching curve by two-time correlation measurements: based on quantum properties of single photon states [21], the absence of coincidence at zero delay gives clear evidence of single photon emission [22]. For several molecules, coincident emission of photons by the different molecules are likely and will result in an autocorrelation function that does not cancel out for zero time-delay. In experiment the effect of background signals makes it impossible that the perfect absence of coincidence at zero delay. Commonly as long as the second-order degree of coherence $g^{(2)}(\tau)$ at zero delay ($\tau=0$) is less than 0.5, which corresponds with two ideal molecules detected, the fluorescence can be considered from a single emitter. However for a small number of blinking molecules, the total number of coincident emission might be too low to reject the single-molecule hypothesis with the antibunching curve.[16] Furthermore the criteria need long time (typically several minutes to tens minutes, depending on the mean photon number $\langle n \rangle$ and the dead-time of detection system) to detect enough photons to statistic the antibunching curve. It is severely affected by the molecule's photostability. And the existence of irreversible photobleaching compelled the long time for single molecules recognition to be insufferable in experiments [23].

Although the phenomenon of photon antibunching is demonstrated mostly clearly by two-time correlation measurements, it is, in principle, also exhibited by the probability P(*n*) that *n* photons are emitted (or detected) in a given time interval *T*. Antibunching implies sub-Poissonian statistics, in the sense that the probability distribution is narrower than a Poisson distribution with the same $\langle n \rangle$ [24]. Traditional photon number counting statistics [25] is seriously affected by the blinking in the fluorescence, due to the molecular triplet state. Then single event photon statistics [20, 26] is suggested to character the single-molecule fluorescence. And based on moment analysis, Mandel's *Q*-parameter, which is defined in terms of the first two photon count moments, is an attractive alternative to two-time correlation measurements. It is quite robust with respect to molecular triplet state effects [27]. In this work we suggest a novel approach to distinguish single-molecule system based on single event photon statistics characterization of single-molecule fluorescence. Using Hanbury Brown and Twiss (HBT) configuration [28], by analyzing and comparing the Mandel parameters of actual single molecule fluorescence and ideal double molecule fluorescence, we present a new criterion based on single event photon statistics measurement.

## II. COMPUTATIONAL METHODS
### A. Detection system of single-molecule fluorescence

Standard confocal microscopy techniques which is the commonly used experimental setup for single-molecule fluorescence detection is summarized in Fig. 1(a) [16], The emitted light is focused on a pinhole in order to reject out-of-focus background light and finally recollimated onto two single photon counting modules (SPCM) after partition by a 50/50 beamsplitter (BS), which is a standard HBT configuration. For each detected photon, the SPCM would generate a TTL pulse. Then a start-stop technique with a time-to-amplitude converter (TAC) allows us to build a coincidences histogram as a function of time delay between two consecutive photodetections on each side of the beamsplitter. The pulses from TAC, whose amplitude is proportional to the time delay, are discriminated according to their heights and accumulated by a multichannel pulse height analyzer (MCA). The second-order degree of coherence at zero delay $g^{(2)}(0)$ can be directly gained and displayed on the screen, as shown in Fig. 1(b) and 1(c). Obviously in Fig. 1(b), $g^{(2)}(0)$ is bigger than 0.5 (dashed line), which indicates that the fluorescence detected is emitted from more than one molecule. In contrast, in Fig. 1(c), $g^{(2)}(0)$ is smaller than 0.5, which indicates that one single molecule was being detected. This method is widely used to distinguish single-molecule in experiments. Nevertheless the process needs a long time to accumulate and display a visible antibunching curve, which is a disadvantage to estimation due to the existence of molecular triplet state. And fluorescent photons should be very weak ($\langle n \rangle$ is less than 0.1), otherwise the start-stop technique will bring an error that can not be ignored.

**B. Single event photon statistics characterization of single-molecule fluorescence**

The single event photon statistics measurement is described in Fig. 2(a). For the influence of detectors' dead time, each SPCM gives only one count (supposedly with 100% detection efficiency) within the dead time for one or more than one incident photon. In order to eliminate the background signals as much as possible, time-gated technique [29] is used after the excitation pulse. The records within the time gates are considered to be the detected signals while all records outside the time gates are rejected. This time-filtering procedure can filter out the real photodetection events from the most of non-synchronous background photocounts not rejected by optical filters, and is often an efficient way to improve the signal-to-noise ratio. The time gate duration must be shorter than the laser period and much longer than the molecule excited-state lifetime so that the probability of discarding a fluorescence signal is negligible. In the above experiments the gate durations are usually ten times the radiative lifetime of the molecule [30]. For a given excitation pulse cycle in detecting time gate which is shorter than the dead time, the number of detected photons cannot exceed two if we using two identical SPCMs operating in the photon counting regime compulsory for our system. The existence of detectors' dead time in each detection channels will results in a non-linear relation-ship between detected photon statistics and source photon statistics. Thus the measured photon probabilities should be corrected.

Fig. 2(b) shows the schematic of our detection setup used for single event photon statistics measurement. The synchronous signals provide a counting time-gate and a nanosecond time delay box is used to compensate the different transport distance between two detection channels. The numbers of pulse cycles which are detected $N_{all}$, in which only one photon is detected $N_{(1)}$, and in which two photons are detected $N_{(2)}$ can be directly gained from a counter. Then the ratio of $N_{(1)}$ to $N_{all}$ and $N_{(2)}$ to $N_{all}$ respectively is the probability of detected one-photon and of detected two-photon in every pulse cycle.

Denoting by $P^{in}(n)$ the photon number probability distribution of incoming light on HBT detection set-up, the non-linear transformation relating this probability to the detected photon probability $P(n = 0, 1, 2)$ is simply computed for 'ideal' detectors. The 'ideal' means that each SPCM clicks with 100% efficiency immediately upon receiving a photon, but that no more than one click can occur in a given repetition period. The joint probability of detecting $i$ photons on SPCM1 and $j$ photons on SPCM2 can be written as $P(i, j)$, $i, j = 0$ or $1$. Actually, there are total four measured photon probabilities: $P(0, 0)$, $P(0, 1)$, $P(1, 0)$ and $P(1, 1)$.

With our experimental detection scheme, random splitting of photons on two sides of 50/50 beamsplitter gives

$$P(0) = P(0,0) = P^{in}(0),$$

$$P(1) = P(0,1) + P(1,0) = \sum_{n \geq 1}^{\infty} P^{in}(n) \frac{1}{2^{n-1}}, \quad (1)$$

$$P(2) = P(1,1) = \sum_{n \geq 2}^{\infty} P^{in}(n) \left(1 - \frac{1}{2^{n-1}}\right).$$

The mean photon number per excitation pulse period is

$$\langle n \rangle = P(1) + 2P(2). \quad (2)$$

Now suppose that in the excitation volume there has more than one molecule, the number is $s$, and all of them are ideal single photon emitters. We call $\eta$ the overall detection efficiency, which includes the optical collection efficiency, all linear propagation losses and quantum efficiency of photon detectors. Then the photon number probability distribution of incoming light on detection set-up is

$$P_I^{in}(n) = \frac{s!}{n!(s-n)!}(1-\eta)^{s-n} \eta^n. \quad (3)$$

Applying Eq. (3) in Eq. (1) one can show

$$P(0) = (1-\eta)^s,$$

$$P(1) = \sum_{n \geq 1}^{s} \frac{s!}{n!(s-n)!}(1-\eta)^{s-n} \eta^n \frac{1}{2^{n-1}}, \quad (4)$$

$$P(2) = \sum_{n \geq 2}^{s} \frac{s!}{n!(s-n)!}(1-\eta)^{s-n} \eta^n \left(1 - \frac{1}{2^{n-1}}\right).$$

Then

$$\langle n \rangle = s\eta(1-\eta)^{s-1} + \sum_{n \geq 2}^{s} \frac{2^n - 1}{2^{n-1}} \frac{s!}{n!(s-n)!}(1-\eta)^{s-n} \eta^n. \quad (5)$$

## C. Mandel's parameter for single event photon statistics

In order to quantify the fluctuations of the number of photons detected per pulse, an important figure of merit is the Mandel parameter $Q = (\langle(\Delta n)^2\rangle - \langle n\rangle)/\langle n\rangle$, where $\langle n\rangle$ is the average number of photons detected within a time interval $T$ and $\langle(\Delta n)^2\rangle$ is the mean variance [31]. The parameter $Q$ is a natural measure of the departure of the variance of the photon number $\langle n\rangle$ from the variance of a Poisson process, for which $Q = 0$. Negative and positive $Q$-values indicate sub- and super-Poissonian behavior, respectively.

From single event photon statistics probability the Mandel parameter can be computed directly

$$Q = \frac{2\mathrm{P}(2)}{\langle n\rangle} - \langle n\rangle. \qquad (6)$$

Note that an ideal single molecule should produce light pulses containing exactly one photon per pulse, triggered with a repetition period, P(2)=0, would yield $Q_i = -\langle n\rangle = -\eta$, which is only limited by the detection efficiency and is the lowest value of the sample's $Q$. Negative $Q$ confirms that single-molecule fluorescence indeed exhibits sub-Poissonian photon statistics which is an explicit feature of a quantum field.

## D. Distinguishing single-molecule using $Q$-parameter

The Mandel's $Q$-parameter provides an alternative for differentiating single-molecule fluorescence system. The $Q$-parameter of ideal double molecules (which is used to model two molecules that can not be separated by diluting) fluorescence can be used as a boundary between that of an actual single molecule fluorescence and actual double molecules fluorescence. From this boundary we can deduce an explicit criterion of single molecules based on photon statistics.

In an experiment a problem we have to face is the background signals, which includes radiation from the environment and the dark counts of the SPCM. Because the scattering background light from surroundings can be thought as a thermal field with very large bandwidth and very short coherent time, the usual photon counting time (nano-seconds) discussed here is much longer than the coherent time and the photocounts of such time-average stationary background show a Poisson distribution. In a dark environment, the SPCM also generates random dark counts that follow a Poisson distribution. Both of these two random counts appear in the Poisson distribution, and thus we can use a weak coherent field with a Poisson photon distribution $P_B^{in}(n) = e^{-\eta\gamma}(\eta\gamma)^n/n!$ with $\gamma$ to simulate the backgrounds. Actual single-molecule fluorescence can be modeled as the superposition of an ideal single molecule and a background emission that can be modeled as a Poisson distribution. Denote by $Q_A$ and $P_A(n)$ the Mandel parameter and single event photon probability statistics of the actual single molecules fluorescence. Using Eq. (1) and Eq. (4), the photon statistics probability can be written as

$$\mathrm{P}_A(0) = e^{-\eta\gamma}(1-\eta),$$

$$P_A(1) = 2\left(e^{-\eta\gamma/2} - e^{-\eta\gamma}\right) + \eta\left(2e^{-\eta\gamma} - e^{-\eta\gamma/2}\right), \qquad (7)$$

$$P_A(2) = \left(1 - e^{-\eta\gamma/2}\right)^2 + \eta\left(e^{-\eta\gamma/2} - e^{-\eta\gamma}\right).$$

Then

$$\langle n \rangle = 2(1 - e^{-\eta\gamma/2}) + \eta e^{-\eta\gamma/2}.$$

The measured mean photon number of fluorescence signals is $S = \eta$, and the measured mean photon number of background signals is $B = 2\left(1 - e^{-\eta\gamma/2}\right)$. According to Eq. (7), we get

$$P_A(0) = (1-S)\left(1 - \frac{B}{2}\right)^2,$$

$$P_A(1) = (S + B - SB)\left(1 - \frac{B}{2}\right), \qquad (8)$$

$$P_A(2) = \frac{BS}{2} + \frac{B^2}{4} - \frac{B^2 S}{4},$$

$$\langle n \rangle = B + S\left(1 - \frac{B}{2}\right).$$

When $P_A(1) \geq 2\sqrt{P_A(2)} - 3P_A(2)$, the signal-to-background ratio (SBR) can be expressed as

$$SBR = \frac{S}{B} = \frac{P_A^2(1)}{2P_A(2)}. \qquad (9)$$

This equation can be directly applied for the measurement of SBR in experiment.

Denote by $Q_D$ and $P_D(n)$ the Mandel parameter and single event photon probability statistics of the ideal double molecules fluorescence. From Eq. (4) one obtains

$$P_D(0) = (1-\eta)^2,$$

$$P_D(1) = 2\eta - \frac{3}{2}\eta^2, \qquad (10)$$

$$P_D(2) = \frac{1}{2}\eta^2,$$

$$\langle n \rangle = 2\eta - \frac{1}{2}\eta^2.$$

Suppose there is no background noise, using Eq. (6) one can show that with the same average photon number $\langle n \rangle$, if $P_A(2) < P_D(2)$, then $Q_A < Q_D$. From Eq. (10) one can obtains

$$P_A(2) < \frac{1}{2}\left(2 - \sqrt{4 - 2\langle n \rangle}\right)^2. \qquad (11)$$

At the same time,

$$P_A(1) > \langle n \rangle - \left(2 - \sqrt{4 - 2\langle n \rangle}\right)^2. \tag{12}$$

Because without background noise it is not possible to have fluorescence from more than one fluorophore in which the $Q$-parameter is less than $Q_D$. So when Eqs. (11) or (12) is satisfied, $Q_A < Q_D$, the fluorescence can be deduced origin from a single molecule system.

### E. Signal-to- background ratio effect on the criterion

However low background signals is a important precondition of the criterion of Eqs. (11) and (12). With the high background signals, $Q_A > Q_D$ will come into existence for a single molecule system. This criterion will not be applicable. So it is necessary to make certain the range of SBR when Eqs. (11) and (12) is satisfied.

Corresponding to Eqs. (11) and (12) using Eqs. (8) and (9) one can obtains

$$SBR_A > SBR_0 = \frac{\sqrt{\langle n \rangle^2 - 2\left(2 - \sqrt{4 - 2\langle n \rangle}\right)^2}}{\langle n \rangle - \sqrt{\langle n \rangle^2 - 2\left(2 - \sqrt{4 - 2\langle n \rangle}\right)^2} - \frac{1}{2}\left(\langle n \rangle - \sqrt{\langle n \rangle^2 - 2\left(2 - \sqrt{4 - 2\langle n \rangle}\right)^2}\right)^2}. \tag{13}$$

It is well known that SBR of the actual single molecule system varies with average photon number $\langle n \rangle$. Based on Eq. (13), the SBR curve versus $\langle n \rangle$ is shown in Fig. 3. When $\langle n \rangle$ is between 0 and 1, the variation range of SBR is within 1.63 to 2.41. So with the same $\langle n \rangle$ when $SBR_A > SBR_0$, for actual single molecule fluorescence and ideal double molecules fluorescence, it has $Q_A < Q_D$. Especially when SBR>2.41, the Eqs. (11) and (12) can be the criterion used to distinguish single molecule system. Contrarily when $SBR_A < SBR_0$, even an actual single molecule system can not satisfy the Eqs. (11) and (12). It is obvious that SBR is also an important parameter of distinguishing single molecule system. Nevertheless most of the SBR in single molecule fluorescence experiments is large than 2.41.

### III. RESULTS AND DISCUSSION

After error correction from appendix one can obtains a more sufficient criterion,

$$P_A(1) > \langle n \rangle - \left(2 - \sqrt{4 - 2\langle n \rangle}\right)^2 - \delta P_A(1) + \langle \Delta P_A(1) \rangle = P_1, \tag{14}$$

$$P_A(2) < \frac{1}{2}\left(2 - \sqrt{4 - 2\langle n \rangle}\right)^2 - \delta P_A(2) - \langle \Delta P_A(2) \rangle = P_2. \tag{15}$$

In the above inequalities the critical values $P_1$, $P_2$ are impacted by three parameters, the mean photon number $\gamma$ of Poissonian background, the mean overall detection efficiency $\eta$, and factor $\Delta$ which represent the unbalance of two channels of imperfect detection system.

Now let us assume that the size of sample cycles $M$ was $10^4$. In this way the statistical fluctuation of the $\langle \Delta P_A \rangle$ induced by finite sample pulse cycles detected can be negligible, which is less than 1/100 of the statistics probability itself. The error $\delta P_A$ caused by imperfect detection system is the main correction factor of the critical value.

Fig. 4 shows the critical values $P_1, P_2$ as the function of mean overall detection efficiency $\eta$ and factor $\Delta = 0.3$, $\gamma = 0.2$, which corresponds to SBR = 5. It is found that the critical values $P_1, P_2$ all increase with increasing efficiency $\eta$ as respected, however the P1 increases slowly while P2 increases fast. Furthermore because in single molecule experiments $P_A(2)$ is less than $P_A(1)$, the effect of the error on $P_2$ is more evident than on $P_1$. So using $P_A(1)$ as the criterion is more feasible.

The experiment data, shown in Table I, is obtained by using typical single molecule experiment setup [20]. The sample were dye molecules Cy5 ($5\times10^{-10}$ M, Molecular Probes) doped in Polymethyl Methacrylate (PMMA, Sigma-Aldrich) polymer films. Molecules in the sample were excited by 50 ps pulses at a wavelength of 635 nm and repetition rate of 2 MHz, generated by a ps pulsed diode laser (PicoQuant, PDL808), at the focus of a confocal inverted microscope (Nikon). Fluorescence photons were detected in two detection channels, using two identical single photon counting modules (SPCMs, PerkinElmer SPCM-AQR-15).

For example, in a typical experiment, to sample 1, during 299613 periods (about 149 ms) yielding 13917 recorded photons including 13902 single photon events, 15 two-photons events. These data allow us to extract the photon probabilities P(0) =0.9535, P(1) = 0.0464, and P(2) =$5\times10^{-5}$ and the mean number of detected photon per pulse $\langle n \rangle$ = 0.0465, would then yield Mandel parameter $Q$= -0.04435.

We observed fluorescence photons for three sample molecules in one polymer film. As shown in Table I, for sample 1 and sample 2, both P(1) are much more than critical values $P_1$ which indicates that a single molecule were being detected; for sample 3, P(1) is less than critical values $P_1$ which indicates that more than one molecule was being detected. As a reference, the interrelated results for coherent light (pulse laser) are attached both from theoretical analysis and experimental measurement. The typical time for single event photon statistics measurement is about 150ms.

## IV. CONCLUSION

We present a fast and robust method to recognise single molecules based on single event photon statistics. Mandel's Q-parameter provides an attractive approach to two-time correlation measurements, because it is easy to implement, requires little time, and is immune with respect to the effects of molecular triplet state. Compared with commonly two-time correlation measurements, our approach has some advantages: (1) the effect of molecular triplet state can be ignored, whereas its effect can only be contained in the non-perfect detection efficiency analysis; (2) ~ms level measurement time is needed as only ~$10^4$ fluorescence photons are needed for photon statistic; (3) it is not limited only to weak photons emitted, which means it is independent with the fluorophores photon intensity. The method can also be applied for the other single emitters recognition, such as single atoms, quantum dots and color centers.

## ACKNOWLEDGEMENTS

This project is supported in part by Natural Science Foundation of China (Grant No.

10674086), 973 Program (Nos.2006CB921603, 2006CB921102), Specialized Research Fund for Doctoral Program of Higher Education of China (2004108002), the Shanxi Provincial Foundation for Leaders of Disciplines in Science, China, the Natural Science Foundation of Shanxi province, China (Grant No. 2007011006), and PCSIRT(IRT0516).
[+] Corresponding to Liantuan Xiao. Email: xlt@sxu.edu.cn.

## APPENDIX: ERROR ESTIMATE

The question of principal importance to ensure the sufficiency of the criterion is error estimation in the single photon statistics detection. As it follows from the detection system, there are two main kinds of errors:

1) A systematic error $P'_A$ caused by imperfect detection system based on HBT configuration. The two sides are unbalanced with the overall efficiency $\eta_1$, $\eta_2$ respectively. Then the photon statistics probability of actual single molecules fluorescence can be written as [26]

$$P'_A(0) = \left(1 - \frac{\eta_1 + \eta_2}{2}\right) e^{-\gamma(\eta_1+\eta_2)/2}, \tag{16}$$

$$P'_A(1) = \left(1 - \frac{\eta_1}{2}\right) e^{-\gamma\eta_1/2} + \left(1 - \frac{\eta_2}{2}\right) e^{-\gamma\eta_2/2} - 2\left(1 - \frac{\eta_1+\eta_2}{2}\right) e^{-\gamma(\eta_1+\eta_2)/2}, \tag{17}$$

$$P'_A(2) = 1 - \left(1 - \frac{\eta_1}{2}\right) e^{-\gamma\eta_1/2} - \left(1 - \frac{\eta_2}{2}\right) e^{-\gamma\eta_2/2} + \left(1 - \frac{\eta_1+\eta_2}{2}\right) e^{-\gamma(\eta_1+\eta_2)/2}. \tag{18}$$

The difference between balanced detection system and unbalanced detection system can be expressed as

$$\delta P_A(1) = P_A(1) - P'_A(1) = \left[2 - \left(1 - \frac{(1+\Delta)\eta}{2}\right) e^{-\Delta\eta\gamma/2} - \left(1 - \frac{(1-\Delta)\eta}{2}\right) e^{\Delta\eta\gamma/2} - \eta\right] e^{-\eta\gamma/2}, \tag{19}$$

$$\delta P_A(2) = P_A(2) - P'_A(2) = \left[\left(1 - \frac{(1+\Delta)\eta}{2}\right) e^{-\Delta\eta\gamma/2} + \left(1 - \frac{(1-\Delta)\eta}{2}\right) e^{\Delta\eta\gamma/2} + \eta - 2\right] e^{-\eta\gamma/2}. \tag{20}$$

In which

$$\eta = \frac{\eta_1 + \eta_2}{2}, \Delta = \frac{\eta_1 - \eta_2}{\eta_1 + \eta_2}. \tag{21}$$

Here $\eta$ is the mean overall detection efficiency and $\Delta$ represent the unbalance of two detection channels. Obviously we can find

$$\delta P_A(1) + \delta P_A(2) = 0. \tag{22}$$

In order to discuss the relative effort of imperfect detection system, normalized relative deviations of $P_A$ can be defined

$$R_1 = \frac{P_A(1) - P'_A(1)}{P_A(1)}, R_2 = \frac{P_A(2) - P'_A(2)}{P_A(2)}. \tag{23}$$

The relative deviation R can also be expressed

$$R_1 = \frac{2-\eta-(1-0.5\eta-0.5\eta\Delta)e^{-\Delta\eta\gamma/2}-(1-0.5\eta+0.5\eta\Delta)e^{\Delta\eta\gamma/2}}{2(1-e^{-3\eta\gamma/2})+\eta(2e^{-3\eta\gamma/2}-1)}, \qquad (24)$$

$$R_2 = \frac{\eta-2+(1-0.5\eta-0.5\eta\Delta)e^{-\Delta\eta\gamma/2}+(1-0.5\eta+0.5\eta\Delta)e^{\Delta\eta\gamma/2}}{\eta-2+e^{\eta\gamma/2}+(1-\eta)e^{-\eta\gamma/2}}. \qquad (25)$$

It is found that the relative deviation $R_1$ is always a minus value, which means the probability $P_A(1)$ of imperfect detection system is more than the balanced one. The imperfect detection system will reduce the probability that photons come into two detectors at the same time from two different channels. The relative deviation $R_1$ decreases with increasing $\eta$ and $\Delta$ as expected, which means that the more difference exists between the two channels, the more departure of $P_A(1)$ exists from that in balanced system.

The relative deviation $R_2$ is nearly independent of mean overall detection efficiency $\eta$ and mean photon number of Poissonian background $\gamma$ with practical ranges of $0<\gamma<1$, and $0<\eta<1$. Be opposite with $R_1$, the relative deviation $R_2$ is always a plus value, the more difference exists between the two channels, the lower $P_A(2)$ will be. And how much $P_A(1)$ increased by imperfect detection system, how much $P_A(2)$ reduced.

2) A random error which is due to the statistical fluctuation of the $P_A$. For every pulse period the fluctuation of the number of the pulses in which there is one photon detected is

$$\langle \Delta N^2 \rangle = \langle N^2 \rangle - \langle N \rangle^2 = P_A(1-P_A). \qquad (26)$$

Now let us assume that M pulse cycles are measured. The fluctuation of the $P_A$ can be estimated that

$$\langle \Delta P_A^2 \rangle = \frac{\langle \Delta N^2 \rangle}{M^2} = \frac{P_A(1-P_A)}{M}. \qquad (27)$$

As evident from Eq. (27), $\langle \Delta P_A^2 \rangle$ has inversely proportional relationship with $M$, which accords with the basic property of statistics. More cycles measured the $P_A$ detected will more close with the real probability.

# REFERENCES

[1] J. Perrin, Ann. Phys. **10**, 133 (1918).
[2] G. Binnig, H. Rohrer, C. Gerber and E. Weibel, Phys. Rev. Lett. **49**, 57 (1982).
[3] G. Binnig, C. F. Quate and C. Gerber, Phys. Rev. Lett. **56**, 930 (1986).
[4] W. E. Moerner and L. Kador, Phys. Rev. Lett. **62**, 2535 (1989).
[5] M.Orrit and J. Bernard, Phys. Rev. Lett. **65**, 2716 (1990).
[6] E. B. Shera, N. K. Seizinger, L. M. Davis, R. A. Keller and S. A. Soper, Chem. Phys. Lett. **174**, 553 (1990).
[7] T. Basché and W. E. Moerner, Nature **355**, 335 (1992).
[8] E. Betzig and R. J. Chichester, Science **262**, 1422 (1993).
[9] S .Nie, D. T. Chiu and R. N. Zare, Science **266**, 1018 (1994).
[10] J. K.Trautman, J. J. Macklin, L. E. Brus and E. Betzig, Sience **369**, 40 (1994).
[11] T. Funatsu, Y. Harada, M. Tokunaga, K. Saito and T. Yanagida, Nature **374**, 555 (1995).
[12] M. Orrit, J. Chem. Phys. **117**, 10938 (2002).
[13] Y. Zhang, C. Gan, and M. Xiao, Phys. Rev. A **73**, 053801 (2006).
[14] J. M. Fernandez, Biophys. J. **89**, 3676 (2005).
[15] A. A. Deniz, M. Dahan, J. R. Grunwell, T. Ha, A. E. Faulhaber, D. S. Chemla, S. Weiss and P. G. Schultz, Proc. Nat. Acad. Sci. USA. **96**, 3670 (1999).
[16] X. Michalet and S. Weiss, C. R. Physique **3**, 619 (2002).
[17] S. Nie and R. N. Zare, Annu. Rev. Biophys.Struct. **26**, 567 (1997).
[18] M. Nirmal, B. O. Dabbousi, M. G. Bawendi, J. J. Macklin, J. K. Trautman, T. D. Harris and L. E. Brus, Nature **383**, 802 (1996).
[19] T. Basché, W. E. Moerner, M. Orrit and H. Talon, Phys. Rev. Lett. **69**, 1516 (1992).
[20] F. Treussart, R. Alléaume, V. Le Floc'h, L. T. Xiao, J.-M. Courty and J.-F. Roch, Phys. Rev. Lett. **89**, 093601 (2002).
[21] P. Grangier, G. Roger and A. Aspect, Eur. Lett. **1**, 173 (1986).
[22] C. Brunel, B. Lounis, P. Tamarat and M. Orrit, Phys. Rev. Lett. **83**, 2722 (1999).
[23] D. L. Kolin , S. Costantino and P. W. Wiseman, Biophys. J. **90**, 628 (2006).
[24] L. Mandel, Opt. Lett. **4**, 205 (1979).
[25] T. Huang, S. L. Dong, X. j. Guo, L. T. Xiao and S. T. Jia, Appl. Phys. Lett. **89**, 061102 (2006).
[26] T. Huang, X. B. Wang, J. H. Shao, X. J. Guo, L. T. Xiao and S. T. Jia, J. Lumin. **124(2)**, 286 (2007).
[27] A. Sanchez-Andres, Y. Chen and J. D. Müller, Biophys. J. **89**, 3531 (2005).
[28] R. Hanbury Brown and R. Q. Twiss, Nature **177**, 27 (1956).
[29] E. B. Shera, N. K. Seizinger, L. M. Davis, R. A. Keller and S. A. Soper, Chem. Phys. Lett. **174**, 553 (1990).
[30] R. Alléaume, F. Treussart, J. M. Courty and J. F. Roch, New J. Phys. **6**, 1 (2004).
[31] R. Short and L. Mandel, Phys. Rev. Lett. **51**, 384 (1983).

TABLE I. Experimental results of single event photon statistics for Dy5 molecules and coherent light. The mean number ⟨n⟩ of detected photons per pulse is calculated by the measured values P(1) and P(2) from Eq. (2). The critical value $P_1$ is estimated from Eq. (14), the relative error is less than 1%. And the $g^{(2)}(0)$ is measured by two-time correlation measurements. Whether the sample detected is a single molecule is determined by that P(1) is more than critical value $P_1$ or not. And the determination is confirmed by the measured value $g^{(2)}(0)$. To a single molecule, the SBR is acquired from Eq. (9).

|  | ⟨n⟩ | P(1) | $P_1$ | $g^{(2)}(0)$ | Single? | SBR |
|---|---|---|---|---|---|---|
| Sample 1 | 0.0465 | 0.0464 | 0.04590 | 0.19 | Yes | 21 |
| Sample 2 | 0.0372 | 0.0370 | 0.03685 | 0.30 | Yes | 7 |
| Sample 3 | 0.0521 | 0.0508 | 0.05150 | 0.65 | No | -- |
| Coherent light (theory) | 0.1046 | 0.0991 | 0.0991 | 1.0 | -- | -- |
| Coherent light (experiment) | 0.1046 | 0.0992 | 0.0995 | 0.98 | -- | -- |

**FIGURE CAPTIONS**

FIG. 1. (a) Schematic of a confocal setup used for single-molecule fluorescence detection. (b) The second-order autocorrelation function measured, which indicates that fluorescence of more than one molecule was being detected. (c) The second-order autocorrelation function measured, which indicates that fluorescence of a single molecule was being detected.

FIG. 2. (a) The single event photon statistics measurement. The sample gate time $\tau_g$, dead time for SPCM $\tau_d$ and laser pulse period $\tau_r$ fulfill $\tau_g < \tau_d < \tau_r$. (b) The schematic of detection setup used for single event photon statistics measurement. The synchronous signals provide a counting time-gate.

FIG. 3. The curve of SBR as a function of the mean photon number $\langle n \rangle$ for actual single molecular photon source. The range out of the shaded portion means the ones that can use the criterion of Eqs. (11) and (12).

FIG. 4. The critical values $P_1$ (solid line), $P_2$ (dashed line) as a function of mean overall detection efficiency $\eta$. And factor $\Delta=0.3$, $\gamma=0.2$, which corresponds to SBR=5. $P_1$ (solid line) using left coordinate, $P_2$ (dashed line) using right coordinate. Curve 2, 3 are the critical values $P_1, P_2$ without error estimate. Curve 1, 4 are the critical values $P_1, P_2$ after error correction.

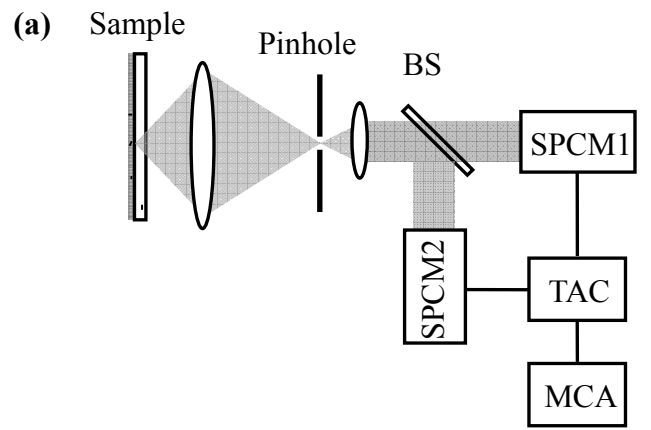

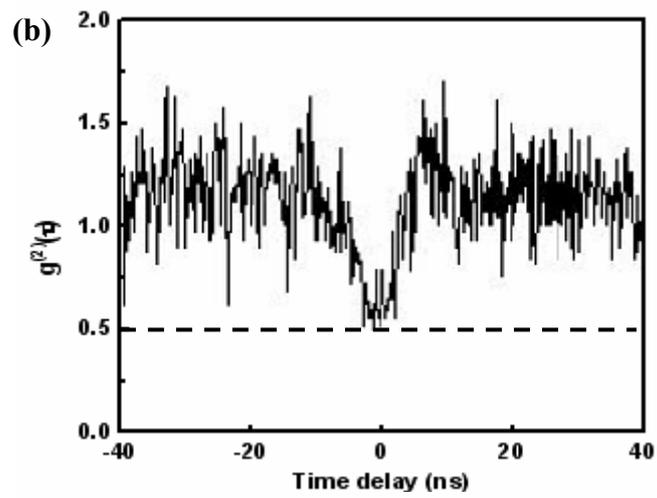

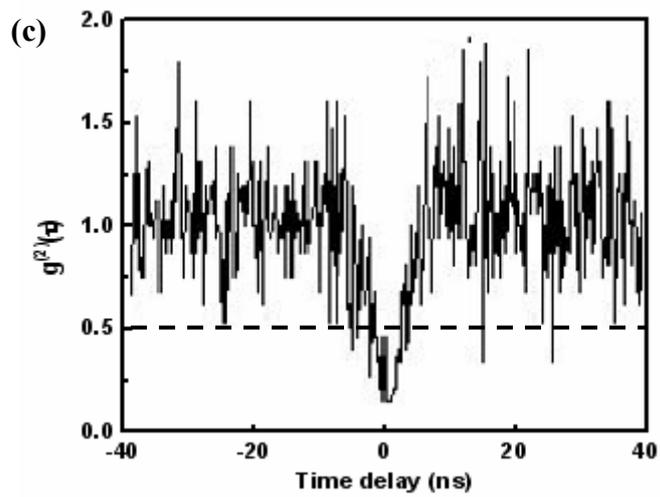

Figure 1

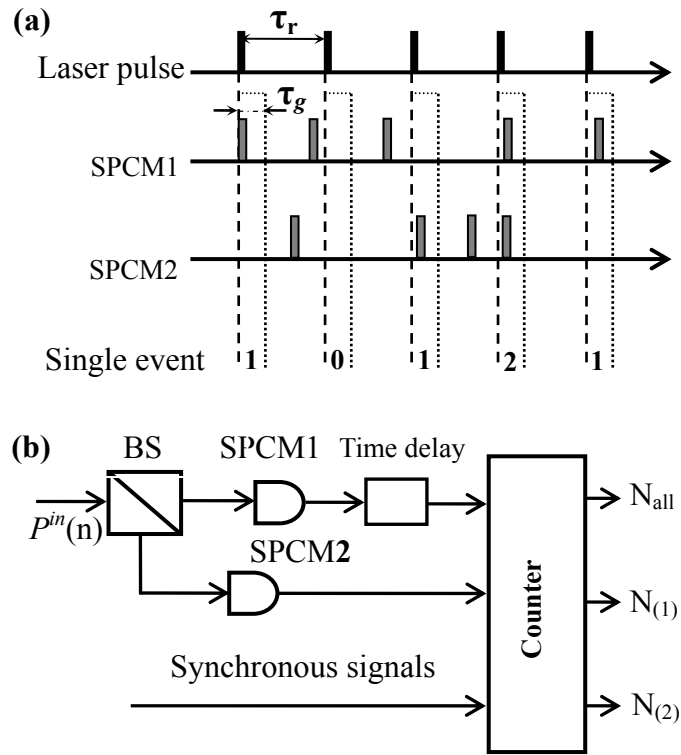

Figure 2

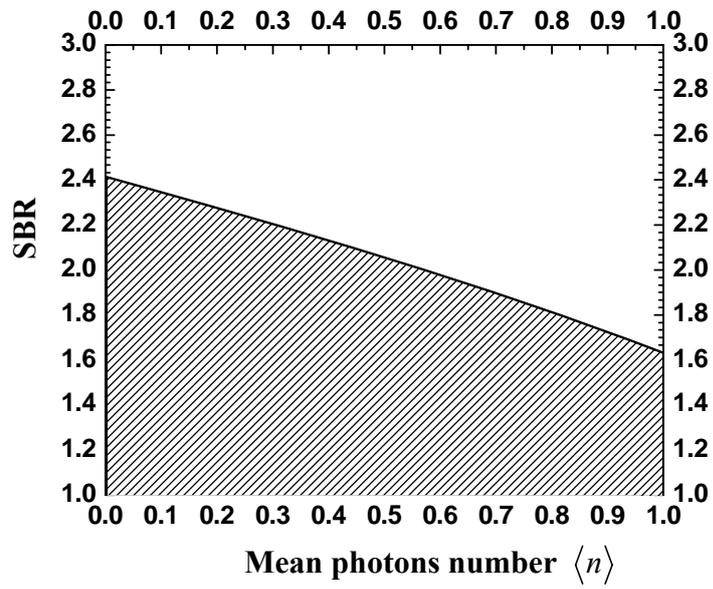

Figure 3

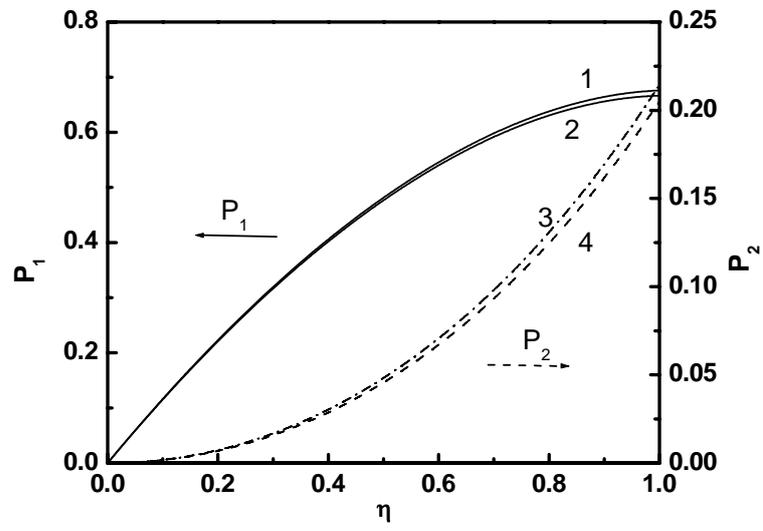

Figure 4